\documentclass[aps,prb,twocolumn,superscriptaddress,amsmath,amssymb]{revtex4}

\usepackage{graphicx}
\usepackage{dcolumn}
\usepackage{bm}

\newcommand{\pdag}{{\phantom{\dagger}}}
\newcommand{\cafeas}{CaFe$_2$As$_2$}
\newcommand{\bafeas}{BaFe$_2$As$_2$}
\newcommand{\srfeas}{SrFe$_2$As$_2$}
\newcommand{\vv}{v_0}

\begin{document}

\title{
Pressure-induced magnetic transition and volume collapse in FeAs superconductors: \\
An orbital-selective Mott scenario
}

\author{Andreas Hackl}
\affiliation{
Institut f\"ur Theoretische Physik, Universit\"at zu K\"oln,
Z\"ulpicher Stra\ss e 77, 50937 K\"oln, Germany
}
\affiliation{
Department of Physics, Harvard University, Cambridge MA 02138, USA
}

\author{Matthias Vojta}
\affiliation{
Institut f\"ur Theoretische Physik, Universit\"at zu K\"oln,
Z\"ulpicher Stra\ss e 77, 50937 K\"oln, Germany
}


\date{\today}

\begin{abstract}
Motivated by pressure experiments on \cafeas,
we propose a scenario based on local-moment physics to explain the simultaneous
disappearance of magnetism, reduction of the unit cell volume, and
decrease in resistivity.
In this scenario, the low-pressure magnetic phase derives from Fe moments, which
become screened in the paramagnetic high-pressure phase.
The quantum phase transition can be described as an orbital-selective Mott
transition, which is rendered first order by coupling to the lattice,
in analogy to a Kondo volume collapse.
Spin-fluctuation driven superconductivity competes with antiferromagnetism and
may be stabilized at low temperatures in the high-pressure phase.
The ideas are illustrated by a suitable mean-field analysis of an
Anderson lattice model.
\end{abstract}


\maketitle


\section{Introduction}
\label{intro}

The discovery of superconductivity in the iron pnictides,\cite{kami08,taka08,yang08}
with the maximum reported transition temperature over 50 K,\cite{ren08}
has spurred intense activities in both experiment and theory.
While the initial research has mainly concentrated on compounds
with $P4/nmm$ structure, e.g. LaFeAsO$_{1-x}$F$_x$,
superconductivity was later also found in oxygen-free materials with $I4/mmm$ structure,\cite{rotter08}
with examples being CaFe$_2$As$_2$ and BaFe$_2$As$_2$.
According to their chemical composition, the two groups have been dubbed
1111 and 122 compounds, respectively.
In both cases, the structure consists of FeAs layers, which are believed to be
responsible for the low-energy electronic properties.

The temperature--doping phase diagram of LaFeAsO$_{1-x}$F$_x$ has been mapped
out in some detail.\cite{buechner08}
At $x=0$ the material undergoes a structural transition into an
orthorhombic phase upon cooling, closely followed by a magnetic
transition. Neutron diffraction shows that the order is of layered antiferromagnetic
type, with an in-plane ordering wavevector $(\pi,0)$.\cite{cruz08}
With increasing $x$, magnetism disappears abruptly at $x\approx4.5\%$ and gives way
to superconductivity, with $T_c$ having little doping dependence up to
$x=0.2$.

This phase diagram suggests interesting similarities to the high-$T_c$ cuprates, where a doping-driven
transition from an antiferromagnet to a superconductor is found as well.
However, a number of differences between the FeAs compounds and the cuprates
are apparent.
(i) The ``undoped'' FeAs materials are not Mott (or charge-transfer) insulators,
but may be characterized as bad metals.
(ii) The FeAs magnetism is not fully consistent with a local-moment picture,
as the size of the ordered moment is unexpectedly small, in the range
$0.25\,\mu_B$ \cite{klauss08} to $0.36\,\mu_B$,\cite{cruz08}
whereas a local moment on Fe is expected to have at least $2\,\mu_B$.
Although magnetic frustration has been identified as a relevant ingredient,\cite{si08}
this alone is not enough to explain the experimental data.
(iii) In the doping regime where superconductivity occurs at low temperatures in the
FeAs materials, the signatures of magnetism appear to be weak,
although the magnetic susceptibility is larger than in a conventional
metal,\cite{klingeler08}
and a magnetic ``resonance'' mode has been reported below $T_c$ in Ba(Fe,Co)$_2$As$_2$.\cite{lumsden08}

From first-principles calculations,\cite{yildirim08}
the most important bands in the iron arsenides
are of Fe 3d character, with sizeable admixtures of As p states.
Importantly, all Fe 3d orbitals appear to contribute to the low-energy properties.
In contrast, in the cuprates only the Cu 3d$_{x^2-y^2}$ orbital (together with O 2p$_{x,y}$)
contributes. Thus, orbital degeneracy and strong hybridization likely play an important
role in the iron arsenides.
(It should be noted that the reported results of first-principles calculations
are partially contradictory, in particular regarding the magnetic properties,
see Ref.~\onlinecite{mazin08}.)

Pressure experiments on \cafeas\ have provided interesting additional
information.\cite{kreyssig,tori,goldman,yu,pratt,canfield}
Starting from the undoped magnetically ordered compound in the orthorhombic phase,
the application of hydrostatic pressure induces a first-order transition at around 0.4 GPa
into a (compressed) tetragonal phase, where magnetism disappears.
Remarkably, there is also a pressure-driven structural transition at
elevated temperatures, from the ambient-pressure tetragonal phase into the compressed
high-pressure tetragonal phase.
This transition is accompanied by a drop in the resistivity, i.e., the
high-pressure phase appears to be a better conductor.\cite{tori}
Band-structure calculations suggest that the disappearance of magnetism is intimately
connected to the pressure-induced volume collapse.\cite{kreyssig,goldman,antropov2}
The high-pressure phase was reported to be superconducting \cite{kreyssig,tori}
with a $T_c$ of about 12\,K over a significant range of pressures, however,
subsequent experiments\cite{yu,canfield} with optimized hydrostatic pressure conditions did not
detect superconductivity down to 4.2\,K, suggesting that the earlier reports were
related to strain-induced superconductivity. Nevertheless, all measurements agree on the
volume collapse and the concomitant loss of magnetism and drop of resistivity.
In addition, superconductivity under high pressures has also been reported\cite{alireza} in
\srfeas\ and \bafeas.

The purpose of this paper is to propose a phenomenological local-moment-based scenario
for the physics of the FeAs compounds, with the primary goal of explaining the pressure
experiments on \cafeas.\cite{kreyssig,tori,goldman,canfield}
Our scenario is based on the assumption that
bands with more itinerant electrons co-exist and interact with more localized (i.e.
moderately to strongly correlated) ones, which renders the problem similar to Kondo or
Anderson models for heavy-fermion metals.
(We note that related ideas on localized and itinerant electrons in iron arsenides have been put
forward in a few recent theory papers\cite{phillips,dai08,kou,dai09} which appeared while this work
was being completed.)
Our analysis will draw analogies to recent developments in heavy-fermion physics,
particular, we shall associate the pressure-induced transition in \cafeas\ with a
variant of the Kondo volume-collapse transition, driven by a large increase of
the hybridization between the itinerant and localized electron bands,
which in turn quenches the magnetism.

The remainder of the paper is organized as follows:
In Sec.~\ref{sec:scen} we sketch our general ideas and point out similarities and
differences between the iron arsenides and classical heavy-fermion materials.
We then illustrate the proposal by a simple mean-field calculation in Sec.~\ref{sec:mf},
which we believe captures important parts of the relevant physics of the \cafeas\ pressure
experiments.
A discussion of implications will wrap up the paper.


\section{Local moments in a correlated Anderson lattice}
\label{sec:scen}

Our proposal is based on the assertion that electronic correlations in the iron arsenides
are sizeable.\cite{si08,phillips,vdb08,laad08}
Then, the magnetism is not purely of weak-coupling type, but instead
local-moment physics is relevant.\cite{foot_itloc}
This view appears consistent with the results of neutron scattering experiments,
which find spin-wave excitations in the antiferromagnetic ground state of \cafeas,
that are well-described by an anisotropic 3d Heisenberg model, at least for long
wavelengths.\cite{mcqueeney}

To explain the fact that the antiferromagnetic phase of FeAs compounds is neither
an insulator nor a good metal, we invoke the existence of (at least) two types of
electrons, one more localized and one more itinerant species. In the spirit of a Kondo or
Anderson lattice model, we shall adopt here the language of ``local-moment'' and
``conduction'' electrons.
In the antiferromagnetic phase, the local-moment electrons carry the magnetism, and the
residual interaction between local-moment and itinerant electrons provides sizeable
scattering, leading to both bad-metal behavior and reduced moment amplitudes.\cite{foot_itloc,mv08}
In the paramagnetic phase, the local-moment electrons are strongly hybridized with the conduction
electrons (i.e. the moments are Kondo-screened), leading to good metallic behavior.
Provided that spin fluctuations are still sizeable,
spin-fluctuation mediated superconductivity can arise at low temperatures.

How does this phenomenological picture tie in with microscopic considerations?
The most plausible scenario is that both types of electrons are primarily of Fe 3d
character, with strong admixtures of As 2p.
As detailed in Ref.~\onlinecite{phillips}, the interplay of p-d hybridization,
spin-orbit coupling, and crystal-field splitting can lead to the two highest occupied
FeAs levels being filled with one electron, but with different correlation strengths.
The electrons in the lower level are more localized, providing a natural basis for our
phenomenological approach.\cite{foot_other}

Assuming the applicability of an Anderson lattice picture, let us further develop our
ideas, concentrating on the transition between the antiferromagnetic and paramagnetic phases.
In fact, many qualitative results can be borrowed from recent theoretical work on
heavy-fermion materials.
There, the zero-temperature transition between a paramagnetic heavy Fermi liquid and an
antiferromagnetic metal can be a standard spin-density-wave transition of a Fermi liquid,
described by a Landau-Ginzburg-Wilson theory of Hertz-Millis type,
or involve the breakdown of Kondo screening.\cite{hvl}
For the latter scenario, a number of theoretical descriptions have been put
forward.\cite{coleman,edmft,senthil,senthil2,pepin}
Within the Kondo-breakdown scenario of Senthil {\em et al.},\cite{senthil2}
the coupling to lattice degrees of freedom has been investigated recently.\cite{hackl08}
If the electron--lattice coupling is sufficiently strong, then the Kondo-breakdown
transition is rendered first order and accompanied by a isostructural volume change,
which is a zero-temperature variant\cite{schmalian} of the classical Kondo volume collapse
transition.\cite{mahan,allen,cyrot,held}
This first-order transition extends to finite temperatures and masks an otherwise
existing magnetic quantum critical point.
(A first-order transition removes the sharp distinction between the Kondo-breakdown and
spin-density-wave transition scenarios.)

In a situation with intermediate correlations, where valence fluctuations are not fully
quenched, it is useful to think about the Kondo breakdown transition as an
orbital-selective Mott transition:\cite{pepin}
In the non-magnetic Fermi liquid phase, the local-moment electrons become itinerant and
strongly hybridized with the conduction electrons, whereas they undergo Mott localization
in the magnetic phase. In fact, such orbital-selective Mott transition have been studied
extensively in two-band Hubbard models.\cite{osmott1}
Based on the cited works, we propose that the pressure-driven transition in \cafeas\ can
be described as orbital-selective Mott transition, which becomes strongly first order
due to electron--lattice coupling. The low-pressure phase displays partial Mott
localization, leading to local-moment magnetism, whereas in the high-pressure phase
strong hybridization quenches the moments and leads to a conventional paramagnetic Fermi
liquid. The pressure-driven iso-structural lattice transition at elevated temperatures
is then the analogue of a Kondo volume collapse.

It is important to point out that our scenario is not in contradiction with
itinerant spin-density-wave descriptions of magnetism in the FeAs compounds. In the Kondo-lattice context,
itinerant and local-moment magnetism can be adiabatically connected (apart from transitions
involving changes of the Fermi surface topology).\cite{mv08}
Therefore, a system at intermediate coupling can in principle be described using both
itinerant and localized electron concepts.
Here, we find it advantageous to employ a strong-coupling language in order to highlight
the nature of the volume collapse transition.
Independent of the language, the driving force of the volume collapse is
the energy gain due to an increasing in the effective hybridization between the bands.

Let us note that there are a few important differences between our
envisioned scenario for the FeAs materials and the phenomenology of heavy-fermions
compounds.
(i) The local-moment electrons in the iron arsenides are probably far from the Kondo
limit. Then valence fluctuations are sizeable, and the picture of Kondo screening does
not literally apply. The coherence temperature on the
high-pressure side of the transition is not small, and the quasiparticles are not very heavy,
in contrast to that of typical heavy fermions.
(ii) The character of the $(\pi,0)$ magnetic order is important to understand the
details of the phase diagram.

The magnetism deserves a few further comments:
In order to explain the magnetic order at wavevector $(\pi,0)$,
both local-moment and itinerant scenarios have been invoked (which are not mutually
exclusive.\cite{foot_itloc,mv08})
In a local-moment picture, the exchange interactions $J_1$ and $J_2$
on the square lattice of Fe atoms (between nearest and next-nearest neighbors, respectively)
have been deduced to be both antiferromagnetic with $J_1\lesssim
J_2$.\cite{si08,yildirim08}
In this regime, the $J_1$-$J_2$ square-lattice Heisenberg antiferromagnet is known to
have a ``layered'' antiferromagnetic ground state with $(\pi,0)$ order.\cite{j1j2a,j1j2b}
In contrast, in an itinerant picture the $(\pi,0)$ order arises from nearly nested Fermi surface
pieces.
Independent of whether the magnetism is better described in an itinerant or localized
picture, magnetic ordering at $(\pi,0)$ in a originally tetragonal environment
breaks the $90^\circ$ lattice rotation symmetry.
This induces an orthorhombic distortion inside the antiferromagnetic phase.
However, the lattice rotation symmetry may also be broken at a higher temperature
than the spin symmetry, in which case the orthorhombic distortion occurs
before the magnetic order.\cite{kiv_nematic,xu08}
This is indeed what happens experimentally,
i.e., magnetic fluctuations are likely the driving force of the structural phase transition.

Last not least, we discuss the possible emergence of superconductivity.
For the Kondo-breakdown scenario, magnetically mediated pairing has been argued to be a
generic instability of the Fermi liquid,\cite{senthil} with the maximum $T_c$
near the continuous Kondo-breakdown transition. In the present situation
of intermediate correlations, the absolute value of $T_c$ will depend strongly on
microscopic details, e.g., of the band structure.
Moreover, if the lattice coupling renders the transition strongly first order,
the system will effectively ``jump'' over the parameter region with large $T_c$.
Thus, while both pressure-induced volume collapse and loss of magnetism are integral and
robust parts of the proposed scenario, superconductivity with a sizeable $T_c$ is more fragile.


\section{Mean-field theory}
\label{sec:mf}

We now illustrate the ideas described in the last section by a
simple model calculation.
We shall refrain from using a realistic band structure with five or more bands,
but instead employ a two-band Anderson lattice model which is
sufficient to capture most of the qualitative physics.

\subsection{Anderson-Heisenberg lattice model}

The starting point of our analysis is an Anderson lattice model,
describing delocalized conduction ($c$) electrons on a lattice which
hybridize with correlated and more localized $f$ electrons on the same lattice.
(Despite the labels $c$ and $f$, both bands may have primarily Fe 3d character.)
To simplify the approximate treatment of magnetism, we supplement our
model by an explicit Heisenberg exchange interaction between the local moments.
The resulting Hamiltonian reads
\begin{eqnarray}
{\cal H} &=&
\sum_{k\sigma} \varepsilon_k c_{k\sigma}^\dagger c_{k\sigma}^\pdag +
\sum_{i\sigma} \epsilon_f^{0} f_{i\sigma}^\dagger f_{i\sigma}^\pdag +
U \sum_i n_{i\uparrow }^f n_{i\downarrow }^f \nonumber\\
&+& \frac{1}{\sqrt{\mathcal{N}}}\sum_{k i\sigma} (V_k e^{-ikR_i} c_{k\sigma}^\dagger f_{i\sigma}^\pdag + H.c.)\nonumber\\
&+& \sum_{ij} J_H(i,j) \vec{S}_i \cdot \vec{S}_{j}
\label{pam}
\end{eqnarray}
in standard notation.
The first term describes conduction electrons with band filling $n_c$,
$\varepsilon_f^0$ ($U$) is the bare $f$ electron energy (Coulomb repulsion),
$V_k$ the hybridization matrix element,
and $\mathcal{N}$ the number of unit cells.
$\vec{S}_i = f_{i\sigma}^\dagger \vec\tau_{\sigma\sigma'} f_{i\sigma'}/2$
is the operator for the spin moment of the $f$ electrons
on site $i$.
Although the system is three-dimensional,
we shall neglect the electronic coupling between the layers for simplicity,
hence the model \eqref{pam} will be treated on a 2d square lattice.
Further we shall assume $V_k\equiv V^0$,
and take the Heisenberg interaction $J_H(i,j)$ to be non-zero
for nearest and next-nearest neighbors, with values $J^0_{1} < J^0_{2}$.
Note that the Kondo limit will not be taken.

\subsection{Lattice distortions and electron--lattice coupling}
\label{sec:distort}

Elastic energy changes of the lattice under application of hydrostatic pressure
depend on the lattice distortion.
For tetragonal lattice symmetry, the most general form of the elastic contribution
to the free enthalpy
is \cite{landau}
\begin{eqnarray}
&&\!\!\!\!G_{\rm lat}(\epsilon_x,\epsilon_y,\epsilon_z)=
\frac{1}{2}\vv c_3 \epsilon_z^2 + \frac{1}{2} \vv c_1 (\epsilon_x^2 +\epsilon_y^2)
+c_{12}\epsilon_x \epsilon_y \nonumber\\
&&~~~~~~~~~~~~~~+ \vv c_{13} \epsilon_z (\epsilon_x +\epsilon_y) +p \vv (\epsilon_x +\epsilon_y
+\epsilon_z).
\label{glat}
\end{eqnarray}
Here, the dimensionless $\epsilon_{x,y,z}$ are the diagonal entries of the strain tensor,
i.e., the relative changes of the lattice parameters of the tetragonal unit cell.
The elastic constants $c_i$, $c_{ij}$, and the reference volume $\vv$ depend on the
material at hand.
Below, we shall choose values for the $c_i$, $c_{ij}$ such that the experimentally
observed lattice distortions of \cafeas\ (Ref.~\onlinecite{kreyssig}) are approximately reproduced.
Importantly, the largest pressure-induced change is the collapse of $\epsilon_z$,
whereas $\epsilon_{x,y}$ even increase slightly at the collapse.

We now construct a model for the electron--lattice coupling, combining
the results from different approaches with simple theoretical arguments.
We start with the strain dependence of the magnetic couplings $J_1$ and $J_2$ which are
responsible for the ambient-pressure magnetism, and will also be relevant for
spin-fluctuation-mediated superconductivity.
In experiment, antiferromagnetic ordering at wavevector
$(\pi,0)$ is energetically stabilized by an orthorhombic distortion with
$\epsilon_x>\epsilon_y$.
Additional evidence is the strong dependence of $J_{1}$ and $J_{2}$ on the strain $\epsilon_z$,
as shown in band structure calculations.\cite{antropov}
In addition, it has been shown\cite{xiang} that changes in the Fe-As bond angle stabilize
antiferromagnetic ordering at wavevector $(\pi,0)$, providing an
additional hint towards the importance of coupling all strain directions
to the exchange couplings.
We therefore parameterize
\begin{eqnarray}
J_{1x}&=&J^0_{1}(1+\gamma_1^\perp \epsilon_x +\gamma_1^z \epsilon_z),\nonumber\\
J_{1y}&=&J^0_{1}(1+\gamma_1^\perp \epsilon_y +\gamma_1^z \epsilon_z),\nonumber\\
J_{2} &=&J^0_{2}\left[1+\gamma_2^\perp (\epsilon_x +\epsilon_y)+\gamma_2^z \epsilon_z\right].
\label{heisen}
\end{eqnarray}
From neutron scattering data and LDA calculations for the orthorhombic phase of
\cafeas\ \cite{mcqueeney} a rough estimate of $\gamma_1^\perp$ can be obtained.
For $S=1/2$, we plug the LDA values $J_{1x}=82$ \,meV and $J_{1y}=20$ \,meV  from Ref.
\onlinecite{mcqueeney} into the parameterization of Eq. \eqref{heisen} and set
$\epsilon_x-\epsilon_y =0.01$, as observed in the orthorhombic phase of
\cafeas.\cite{kreyssig} This calculation yields the estimate $\gamma_1^\perp \approx 70$.
The size and sign
of $\gamma_1^\perp$ might well account for the observed orthorhombic distortion and
the $(\pi,0)$ ordering vector, as argued in Ref. \onlinecite{yildirim08}.
From band-structure calculations for LiFeAs \cite{antropov} it is suggested that $\gamma_1^z$ and $\gamma_2^z$
are numbers of $O(10)$, whereas their signs turn out to depend on pressure. To account for
the vanishing superconductivity at high pressures, we chose their sign to be positive,
since $\gamma_1^z$ and $\gamma_2^z$ dominate the changes in $J_{1}$ and $J_{2}$
at high pressures.\cite{gammafoot}
Finally, $\gamma_2^\perp$ is of subleading influence within our
calculation as long as it is of $O(10)$ or smaller, which is suggested by the order of
magnitude of $\gamma_1^z$ and $\gamma_2^z$.

A crucial ingredient for our model is the strain dependence of the hybridization,
which we assume to be the dominant mechanism to drive the volume collapse, as observed
in the Kondo volume collapse in Cerium and other materials.
In general, the mechanism for the Kondo volume collapse transition is a gain in
hybridization energy via a structural distortion. 
A plausible parameterization is
\begin{equation}
V = V^0 \left[1+ \gamma^\perp (\epsilon_x +\epsilon_y)+\gamma^z \epsilon_z\right].
\label{hybrid}
\end{equation}
As our model is to be understood as an effective model, $V$ is related to actual
band structure parameters in a likely complicated fashion, and
information about its strain dependence is not available.
From the Kondo volume collapse model it is known \cite{allen,schmalian}
that $\gamma^\perp$, $\gamma^z$ should be chosen of $O(1)$ to reproduce volume collapses
of $O(10\,\%)$.
Given the experimental behavior of the \cafeas\ lattice constants,\cite{kreyssig}
$\gamma^z<0$ is crucial. We shall also take $\gamma^\perp>0$, but this is subdominant.

Finally, we neglect any dependence of the $c$ dispersion $\varepsilon_k$ and the $f$
energy $\epsilon_0$ on lattice strain.
While such a dependence certainly exists, it will not qualitatively change our picture,
which is dominated by the effects parameterized in equations (\ref{heisen}) and (\ref{hybrid});
moreover, a pressure dependence of the bandwidth can be absorbed in a pressure dependence of
the reference energy scale of our calculation.

\subsection{Fermionic mean-field theory}
\label{sec:mfeq}

The model \eqref{pam} will be solved using a standard mean-field theory,
with a fermionic representation of the local moments and slave bosons to deal with
strong local repulsions. Theories of this type have been extensively used in the study
of the Kondo and Anderson lattice and also allows to deal with the magnetic exchange
term in Eq. \eqref{pam}.\cite{burdin,coqblin,senthil2,pepin}

In the limit of infinite Coulomb repulsion $U$, the physical electron can be represented by
a spinless boson $r_i$ and an auxiliary fermion $\bar{f}_{i\sigma}$,
$f_{i\sigma}=r_i^\dagger\bar{f}_{i\sigma}$,
together with the constraint
\begin{equation}
r_i^\dagger r_i^\pdag + \bar{f}_{i\sigma}^\dagger \bar{f}_{i\sigma}^\pdag =1 .
\end{equation}
The auxiliary bosons $r_i$ will be treated on the mean-field level.
The Heisenberg part of the Hamiltonian, represented in pseudo-fermions $\bar{f}_{i\sigma}$,
has to be decoupled.
Guided by the presence of potential magnetic and superconducting instabilities,
we introduce mean fields according to
\begin{eqnarray}
\vec{M}_r  &=& \frac{1}{2} \langle \bar{f}_{r\sigma}^\dagger \vec{\tau}_{\sigma \sigma^\prime} \bar{f}_{r\sigma^\prime}^\pdag \rangle, \nonumber\\
\Delta_{ij}&=& -\langle\bar{f}_{i\uparrow} \bar{f}_{j\downarrow} -\bar{f}_{i\downarrow} \bar{f}_{j\uparrow}\rangle.
\end{eqnarray}
The decoupling of the quartic interaction takes the form (additional constants omitted):
\begin{eqnarray}
\vec{S}_i \cdot \vec{S}_{j} &=&
x \left[\frac{1}{4}\vec{M}_i \cdot \bar{f}_{j\sigma}^\dagger \vec\tau_{\sigma\sigma'} \bar{f}_{j\sigma'} +
(i\leftrightarrow j) \right] \nonumber\\
&+& (1-x) \left[\frac{1}{2}\Delta_{ij} (\bar{f}^\dagger_{i\uparrow} \bar{f}^\dagger_{j\downarrow} -\bar{f}^\dagger_{i\downarrow}
\bar{f}^\dagger_{j\uparrow} ) + h.c. \right].
\end{eqnarray}
We have introduced an additional decoupling parameter $x \in (0,1)$,
where $x=0$ corresponds to an Sp($N$) large-$N$ limit \cite{senthil} and $x=1/2$ to
unrestricted Hartree-Fock. Physically, $x$ regulates the balance between ordered
magnetism and superconductivity, and we shall choose $x=0.3$ and 0.4 below.

It remains to specify the spatial dependence of the mean-field parameters.
The obvious parameterization for the
magnetization $\vec{M}_r$  is $\vec{M}_r=m_s \exp(i\vec{Q}\cdot\vec{r})$ with
$\vec{Q}=(\pi,0)$.\cite{saddlefoot}
The complex pairing fields $\Delta_{ij}$ live on the nearest-neighbor and next-nearest-neighbor
bonds of the square lattice. In the general orthorhombic case, we focus on saddle
points with $\Delta_x$ on horizontal bonds, $-\Delta_y$ on vertical bonds,
and $\pm \Delta_{xy}$ on diagonal bonds,
such that the pairing term in $k$-space reads
$\Delta_{\vec k} = \Delta_x \cos k_x - \Delta_y \cos k_y  + 2\Delta_{xy} \sin k_x
\sin k_y $.
For a tetragonal lattice, $\Delta_x = \Delta_y$, and the pairing is a mixture of
$d_{x^2-y^2}$ and $d_{xy}$ symmetric terms.\cite{saddlefoot}
Finally, the slave boson is chosen to be uniform, $r_i\equiv r_0$.

The mean-field amplitudes are obtained from the saddle-point equations
\begin{eqnarray}
\bar{\Delta}_{x} &=& \frac{2}{\mathcal{N}}  \sum_k \langle \bar{f}_{k\uparrow}^\dagger  \bar{f}_{-k\downarrow}^\dagger \rangle \cos k_x  \nonumber\\
\bar{\Delta}_{y} &=& -\frac{2}{\mathcal{N}} \sum_k \langle \bar{f}_{k\uparrow}^\dagger  \bar{f}_{-k\downarrow}^\dagger \rangle \cos k_y  \nonumber\\
\bar{\Delta}_{xy}&=& \frac{1}{\mathcal{N}}  \sum_k \langle \bar{f}_{k\uparrow}^\dagger  \bar{f}_{-k\downarrow}^\dagger \rangle
2\sin k_x \sin k_y  \nonumber\\
m_s &=& \frac{1}{4\mathcal{N}} \sum_k \langle  \bar{f}_{k\uparrow}^\dagger \bar{f}_{k+Q\uparrow} - \bar{f}_{k\downarrow}^\dagger \bar{f}_{k+Q\downarrow}^\pdag + h.c.\rangle \nonumber\\
1-r_0^2&=& \frac{1}{\mathcal{N}}  \sum_{k\sigma} \langle \bar{f}_{k\sigma}^\dagger \bar{f}_{k\sigma}^\pdag\rangle ,
\end{eqnarray}
where $\bar{\Delta}$ denotes the complex conjugate of ${\Delta}$.

\subsection{Phases and electronic phase diagram}

The electronic mean-field theory specified above can display the following phases
(not all of which will appear for our choice of parameters):
\begin{itemize}
\item Decoupled, with $r_0=0$, $m_s=0$, $\Delta=0$, describing a paramagnetic
high-temperature regime with localized $f$ electrons.
\item Fractionalized Fermi liquid (FL$^\ast$,) with $r_0=0$, $m_s=0$, $\Delta\neq 0$.
This is a paramagnetic phase, where conduction electrons alone form a ``small'' Fermi surface
(FS) and are decoupled from a fractionalized spin liquid of paired spinons.
This phase was introduced in Ref.~\onlinecite{senthil}, but will not play a role here.
\item Local-moment antiferromagnet (AFM), with $r_0=0$, $m_s\neq 0$. Here, $\Delta$ may be zero
or finite, the latter case reflecting residual spinon pairing.
\item Paramagnetic Fermi liquid (FL), with $r_0\neq0$, $m_s=0$, $\Delta=0$ with itinerant $f$
electrons and a ``large'' Fermi surface.
\item Antiferromagnetic Fermi liquid, with $r_0\neq0$, $m_s\neq0$, $\Delta=0$ with
itinerant $f$ electrons.
\item Paramagnetic superconductor (SC), with $r_0\neq0$, $m_s=0$, $\Delta\neq0$, obtained from
pairing in the large-FS Fermi liquid.
\item Antiferromagnetic superconductor, with $r_0\neq0$, $m_s\neq0$, $\Delta\neq0$.
\end{itemize}
A zero-temperature transition from a phase with $r_0\neq0$ to $r_0=0$
is associated with Mott localization of the $f$ electrons, i.e.,
an orbital-selective Mott transition.
A continuous $T\!=\!0$ transition of this type will survive beyond mean-field in
the paramagnetic case,\cite{senthil,senthil2}
whereas it becomes a crossover in the antiferromagnetic case.\cite{mv08}

\begin{figure}[t]
\includegraphics[clip=true,width=6.5cm]{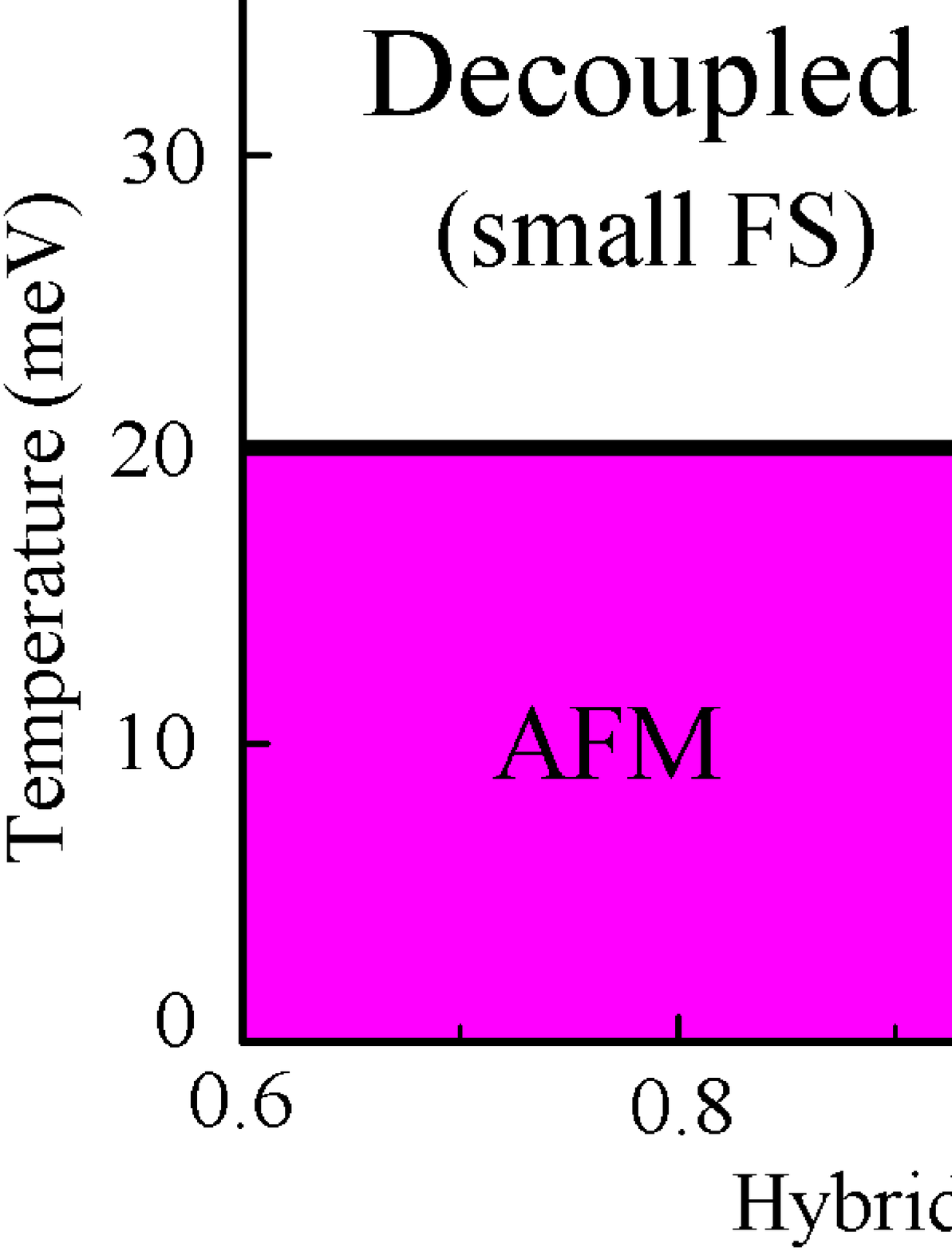}\\
\vspace*{0.4cm}
\includegraphics[clip=true,width=6.5cm]{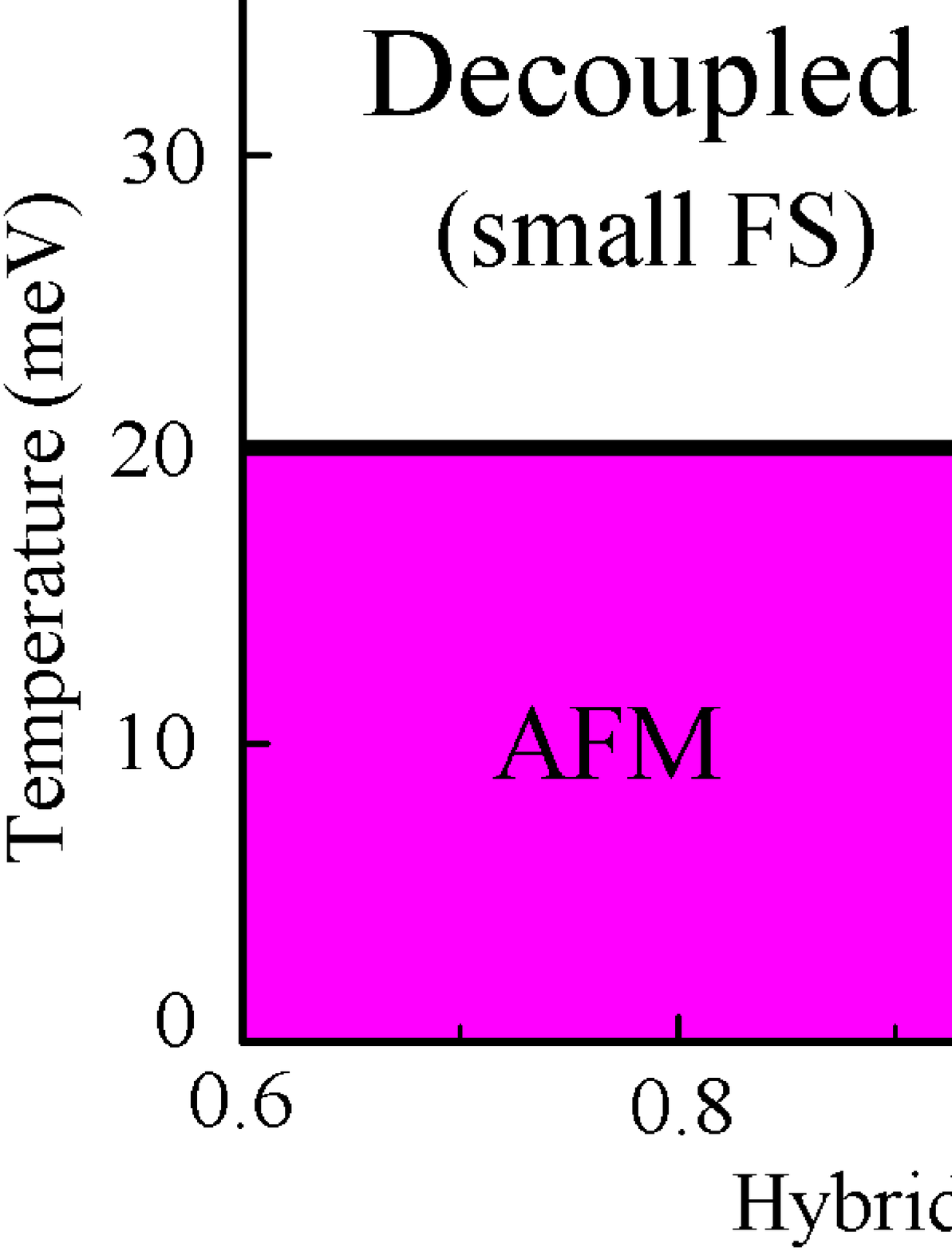}
\caption{
\label{plain}
Electronic mean-field phase diagram in the temperature--hybridization plane,
for fixed lattice parameters $\epsilon_{x,y,z}=0$.
Top: Mean-field decoupling parameter $x=0.3$. Bottom: $x=0.4$.
Thick (thin) lines denote first-order (continuous) phase transitions.
For small hybridization, the $f$ electrons are localized, and magnetism dominates at low
$T$, whereas large hybridization leads to itinerant $f$ electrons and superconductivity.
For details and parameters see text.
}
\end{figure}

For the numerical calculations, serving as an illustration of our ideas from Sec.~\ref{sec:scen},
we choose parameters as follows:
The $c$ electron dispersion consists of nearest-neighbor hopping,
$\varepsilon_k= -2 t (\cos k_x +\cos k_y)$, with $t=0.5$\,eV and filling $n_c=0.8$.
The resulting band width is comparable to the energy range where the Fe DOS is
sizeable.\cite{haule}
The $f$ level position is set slightly below the lower band edge to obtain moderate
valence fluctuations, $\epsilon_f=-2.3$\,eV.
The properties of the local-moment antiferromagnetic phase are
determined by the exchange couplings $xJ_1$ and $xJ_2$.
For a stable AFM ground state with wavevector $(\pi,0)$, we choose the ratio
$J_2^0/J_1^0 = 1.5$.
We employ two parameter sets: $x=0.3$, $J_1^0=200$\,meV, $J_2^0=300$\,meV
and $x=0.4$, $J_1^0=150$\,meV, $J_2^0=225$\,meV.
The resulting $xJ_1^0$ and $xJ_2^0$ coincide with the theoretical values
for \bafeas\ from Ref.~\onlinecite{andersen}. For \cafeas, experimental
and theoretical values have been determined in the orthorhombic phase in
Ref.~\onlinecite{mcqueeney}. For our model with $S=1/2$, those results
yield  $J_1 \approx (J_{1x}+J_{1y})/2 \approx 40$\,meV and $J_2\approx 50$\,meV.

The electronic phase diagram, obtained for fixed $\epsilon_{x,y,z}=0$
as function of temperature $T$ and hybridization $V_0$, is shown in Fig. \ref{plain},
for decoupling parameters $x=0.3$ and 0.4.
We find magnetism and superconductivity to be mutually exclusive and separated by a
first-order transition:
At this transition, both $m_s$ and $r_0$ jump, i.e., the system switches from a
local-moment dominated antiferromagnet to a fully itinerant superconductor.
At low $T$, spinon pairing co-exists with local-moment antiferromagnetism on the
small-$V$ side of the phase diagram.
The thermal magnetic transition is very weakly first order within our accuracy,
but we cannot exclude it to be continuous.
The superconductivity is of $d_{xy}$ character, i.e. driven by the exchange interaction $J_2$.
(A small $id_{x^2-y^2}$ admixture develops at low temperatures for $x=0.3$.)
The finite-temperature transition between the decoupled regime and the FL will be smeared
into a crossover by fluctuations beyond mean field; the other transitions are accompanied
by physical symmetry breaking and survive.

A variation of the decoupling parameter $x$ within the range $x\in (0.25,0.45)$
(keeping $xJ_{1,2}^0$ fixed)
mainly influences the stability of the superconducting phase, Fig.~\ref{plain}.
Increasing $x$ disfavors pairing and further stabilizes antiferromagnetism.
For small $x<0.25$ a paramagnetic FL$^\ast$ phase is realized at small hybridization $V_0$,
whereas for large $x>0.45$ superconductivity disappears. For $x\geq 0.5$,
antiferromagnetism appears even in the large-FS Fermi-liquid regime.
(A related mean-field theory with spinon hopping (instead of pairing)
generically displays an itinerant antiferromagnetic phase.\cite{senthil2})
In general and beyond simple mean-field approximations, the superconducting $T_c$ will
sensitively depend on band structure details and nesting conditions;
a variation of $x$ in our calculations mimics such changes.

\subsection{Phase diagram with electron--lattice coupling}

\begin{figure}[t]
\includegraphics[clip=true,width=6.5cm]{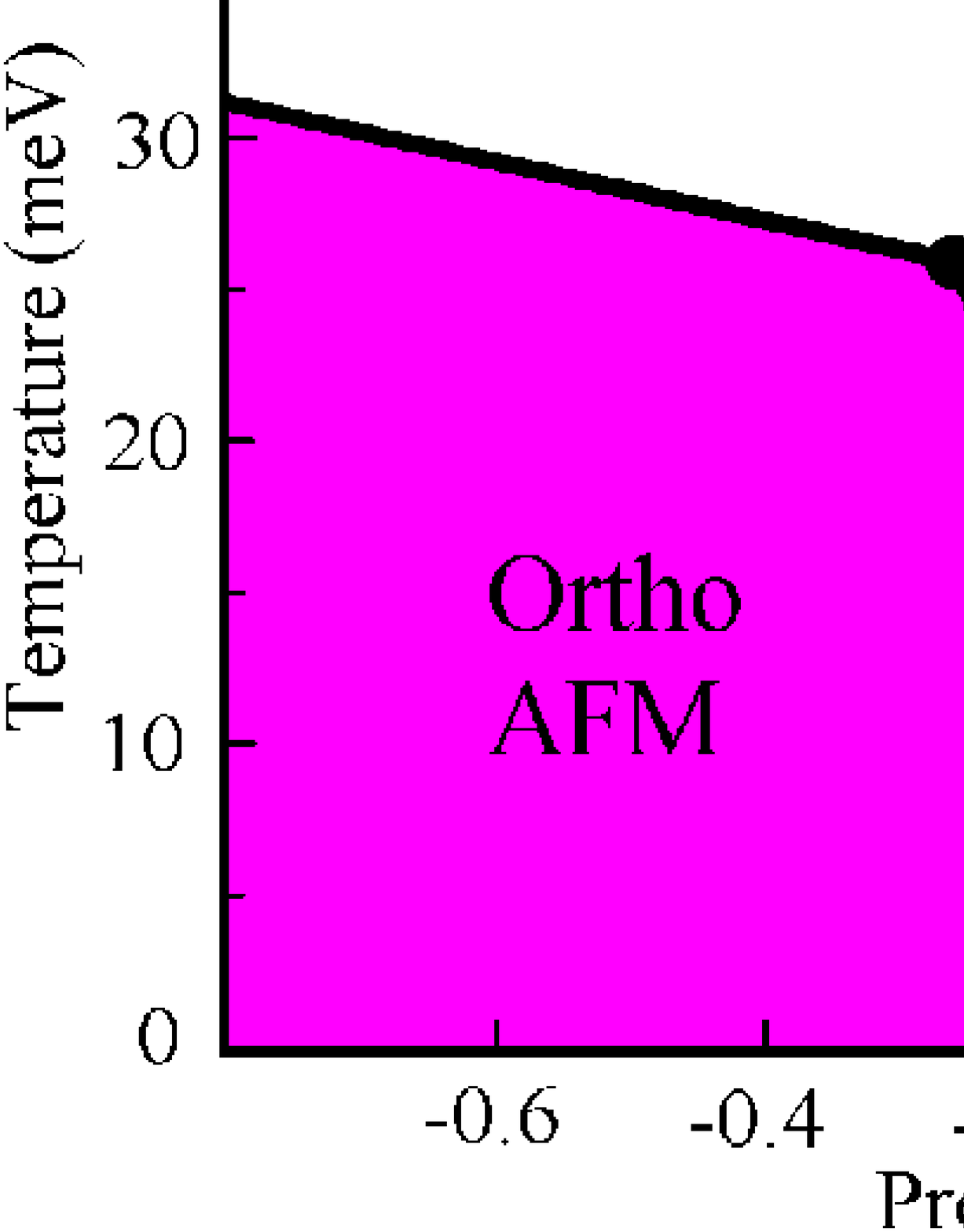}\\
\vspace*{0.4cm}
\includegraphics[clip=true,width=6.5cm]{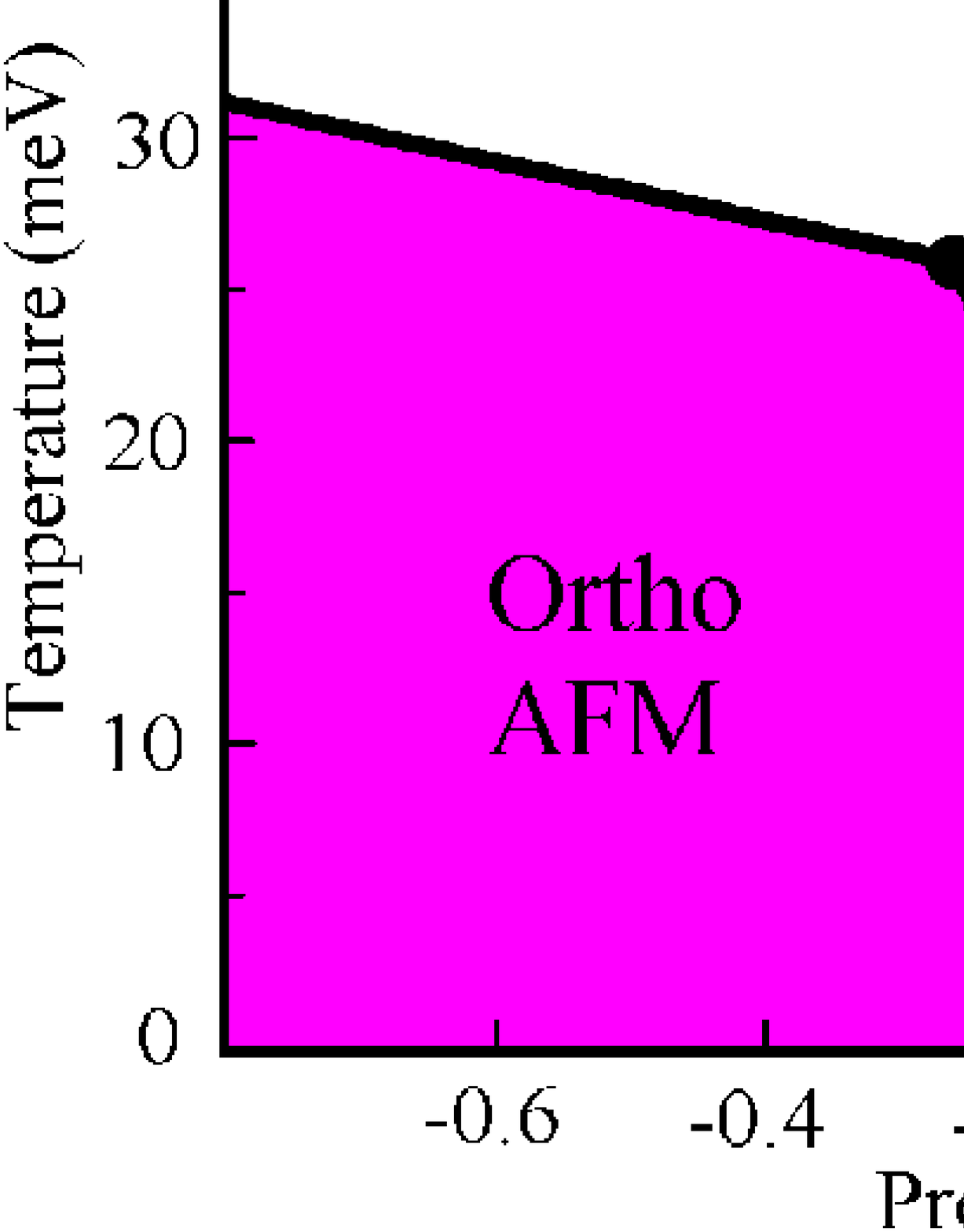}\\
\caption{
\label{collapsephase}
Mean-field phase diagram including electron--lattice coupling in the temperature--pressure plane.
Top: Mean-field decoupling parameter $x=0.3$. Bottom: $x=0.4$.
Thick (thin) lines denote first-order (continuous) phase transitions.
The isostructural volume-collapse transition displays a critical end point
at a temperature of roughly 200\,meV.
See text for details.
}
\end{figure}

The central result of our mean-field study is the phase diagram
in Fig. \ref{collapsephase}, which accounts for lattice distortions and external
pressure.
It has been obtained from minimizing the free enthalpy $G_{\rm el}+G_{\rm lat}$,
where $G_{\rm lat}$ is in Eq.~\eqref{glat} and
$G_{\rm el}$ is the electronic contribution according to the Hamiltonian ${\cal H}$
\eqref{pam}, with the lattice dependence of all parameters as in Sec.~\ref{sec:distort}
and the mean-field approximation as in Sec.~\ref{sec:mfeq}.
As experimental data for the elastic constants of the $122$ materials were not
available to us, we choose elastic constants of $c_1=441$\,kBar, $c_3=198$\,kBar, and
$c_{12}=c_{13}=66$\,kBar.
The employed electron--lattice couplings are
$\gamma^\perp = 3.1$, $\gamma^z = -5.2$, $\gamma_1^\perp = 35$, $\gamma_1^z =1.0$,
$\gamma_2^\perp = 8.0$, and $\gamma_2^z = 5.0$, see Sec.~\ref{sec:distort}.
The values for $c_i$, $c_{ij}$, $\gamma^\perp$, and $\gamma^z$ were adjusted to reproduce the
experimentally observed lattice distortions. The couplings $\gamma_1^z$,
$\gamma_2^\perp$ and $\gamma_2^z$ had little influence on
the lattice distortions.
They were adjusted to reproduce the experimentally
observed phase boundary of the superconducting phase.

\begin{figure}
\includegraphics[clip=true,width=3.9cm]{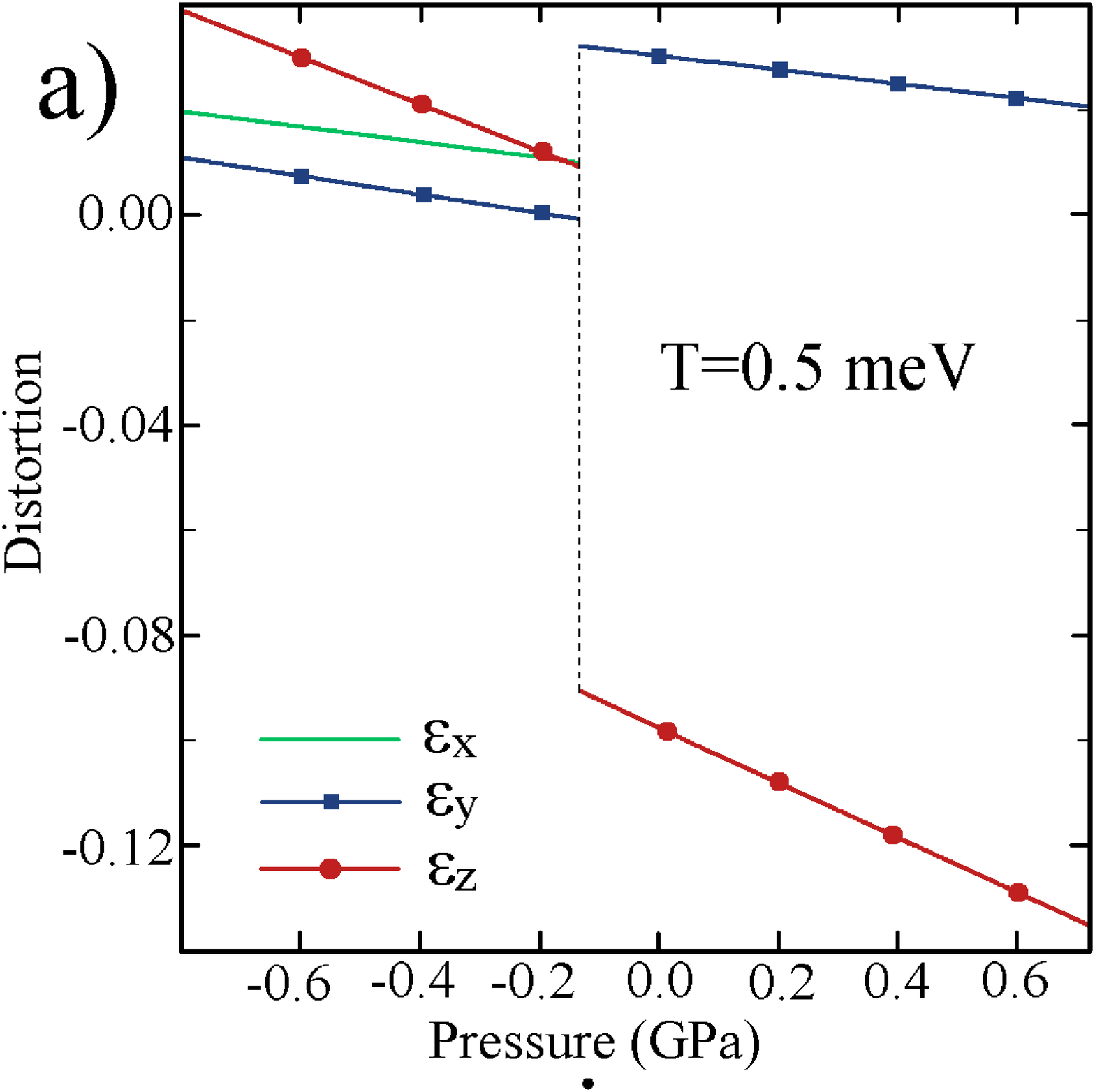}\hspace*{10pt}
\includegraphics[clip=true,width=3.9cm]{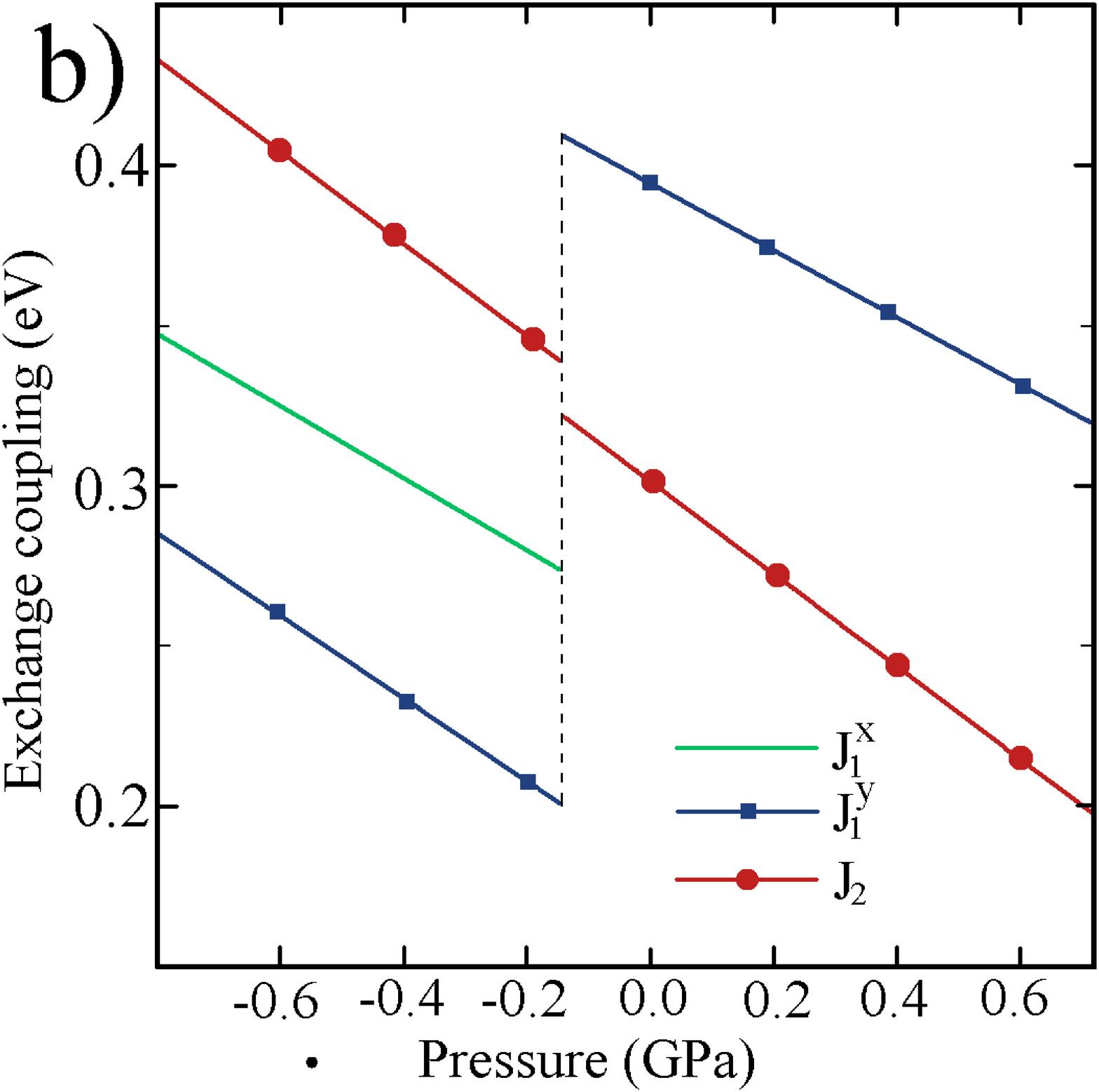}\\[5pt]
\includegraphics[clip=true,width=3.9cm]{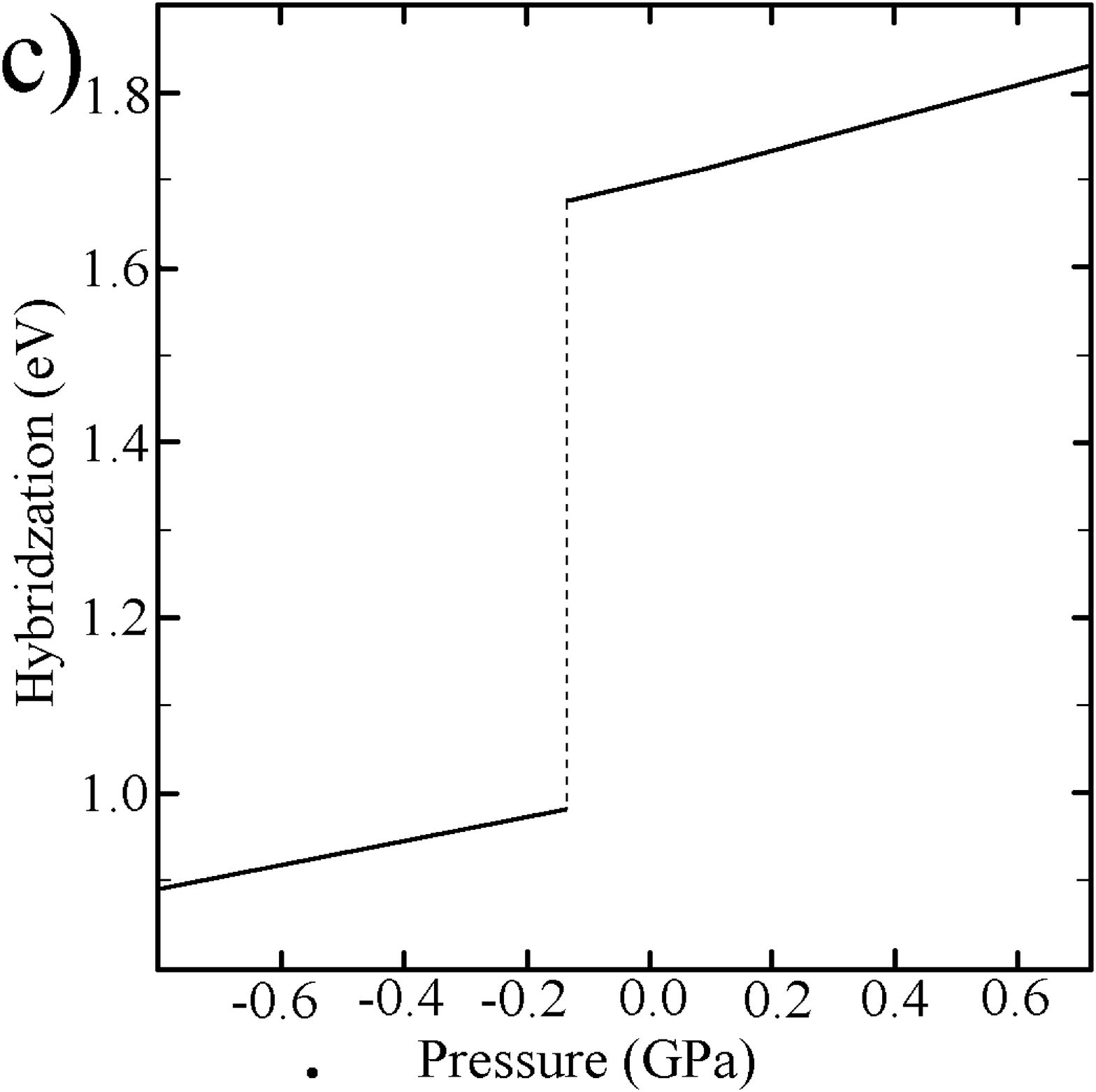}\hspace*{10pt}
\includegraphics[clip=true,width=3.9cm]{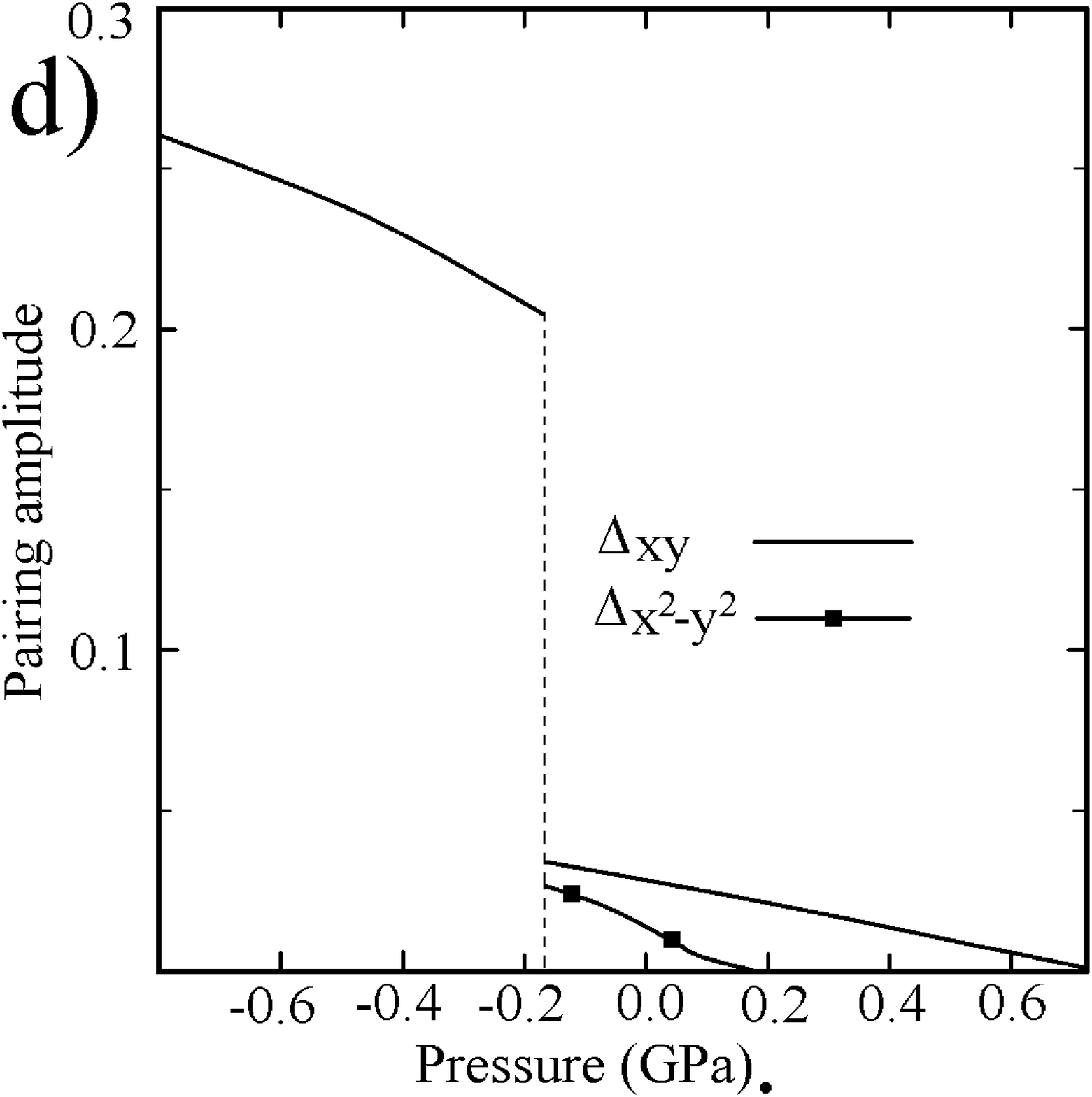}
\caption{
\label{parameters}
Pressure dependence of several microscopic parameters for low $T=0.5$\,meV and
decoupling parameter $x=0.3$.
a) Lattice parameters $\epsilon_{x,y,z}$. Below temperatures of 30\,meV,
the $\epsilon$ depend only weakly on temperature, therefore this plot is representative
for most parts of the phase diagram.
b) Magnetic exchange constants $J_{1,2}$.
c) Hybridization $V$.
d) Pairing fields $\Delta$.
Note that $\Delta$ reflects superconductivity only in the Fermi-liquid regime
at large pressure, whereas it only describes spinon pairing
in the orbital-selective Mott phase at small pressure.
}
\end{figure}

With this parameter choice, the main effect of increasing pressure is an increasing
hybridization, while the magnetic exchange couplings decrease somewhat.
Consequently, the topology of the phase diagram is similar to that of Fig.~\ref{plain},
however, with a few crucial differences:
(i) The orbital-selective Mott transition becomes strongly first order,
at both low and high temperatures.
(ii) The antiferromagnetic phase is accompanied by an orthorhombic lattice distortion,
and the thermal phase transition is of first order.
(iii) Pressure drives a volume collapse transition, which is tetragonal $\leftrightarrow$ tetragonal
at elevated $T$ and orthorhombic $\leftrightarrow$ tetragonal at low $T$.
The pressure dependence of various microscopic parameters is shown in Fig.~\ref{parameters}
for low temperatures.

Our results reproduce salient features of \cafeas\ as reported in
Refs.~\onlinecite{kreyssig,canfield}. In particular, the phase diagram
Fig.~\ref{collapsephase} is very similar to Fig.~3c of Ref.~\onlinecite{kreyssig}.
For $x=0.3$, our superconducting transition temperature close to the critical pressure
is of order 50\,K, whereas the N\'eel temperature is somewhat below room temperature.
While these values could be brought closer to the experimental ones of Ref.~\onlinecite{kreyssig},
we refrain from doing so, because inter-plane coupling and fluctuation effects, both absent from
our treatment, will significantly modify the mean-field result.
For $x=0.4$ the pairing fields are much smaller and the range of superconductivity is
strongly reduced; for $x=0.45$ (not shown) $T_c$ is below 0.1\,meV,
broadly consistent with the more recent experiments of Ref.~\onlinecite{yu}.
All other parameters behave essentially identically as compared to $x=0.3$.
Hence, we conclude that the volume collapse and orbital-selective Mott transition
are robust features of our model, but, in contrast, superconductivity with a sizeable $T_c$
is not.

The present mean-field approach is not able to distinguish the nematic transition from the
magnetic transition, which are separated in experiment, but coincide in
Fig.~\ref{collapsephase}. This deficiency could be in principle repaired by including an
additional nematic order parameter in the mean-field treatment, but would not change the
other features of the phase diagram.

The quantitative values of the lattice distortions in Fig.~\ref{parameters}a are close
to the experimental ones. The pressure dependence of the magnetic exchange,
Fig.~\ref{parameters}b, is relatively strong and not confirmed experimentally,
although the qualitative behavior is suggested by neutron scattering data and band
structure calculations.
In our calculation, the pressure dependence of $J_{1,2}$ is required to create a pressure
dependence of both the Neel temperature and the superconducting $T_c$ comparable to experiment.
(In principle, it is conceivable that fluctuation effects not captured by our treatment
have a similar effect.)
The hybridization $V$ displays a large jump at the critical pressure, Fig.~\ref{parameters}c --
this is, within our theory, the mechanism which balances the increase in elastic energy connected
with the volume collapse.
The pairing fields undergo several changes as function of pressure,
see Fig.~\ref{parameters}d for $x=0.3$,
which depend sensitively on the pressure dependence of the exchange couplings
and other microscopic parameters.
The dominant pairing is of $d_{xy}$ symmetry as above, both for the spinon pairing
of localized $f$ electrons in the magnetic small-pressure phase
as well as for the superconductivity of itinerant $f$ electrons at high pressure
(again with a small $id_{x^2-y^2}$ admixture at low temperatures, which disappears for
$x=0.4$).
Slight changes of $J_2/J_1$ stabilize saddle points with other pairing symmetries,
but leave the non-superconducting part of the phase diagram essentially unchanged.
Finally, the $f$ occupation jumps from 1 to $\approx 0.80$ across the transition for all
temperatures below 60\,meV (not shown).


\section{Discussion}
\label{disc}

In this paper, we have proposed a theoretical scenario to rationalize the
pressure-induced phase transitions in \cafeas.
Underlying the scenario is the physics of the Anderson lattice,
used for heavy-fermion metals:
Strongly localized electrons hybridize with more itinerant electrons.
Microscopically, both may be primarily of Fe 3d character, as discussed in
Ref.~\onlinecite{phillips} (although this is not required within our phenomenological
approach).
At ambient pressure, the localized electrons order antiferromagnetically in a collinear
arrangement at low temperatures, accompanied by an orthorhombic lattice distortion.
The system is weakly metallic, due to the presence of the itinerant carriers
with a ``small'' Fermi volume.
Increasing pressure drives a transition towards a paramagnetic Fermi liquid, where the
previously localized electrons become itinerant and non-magnetic.
The coupling to the lattice degrees of freedom in \cafeas\ renders this transition strongly first order --
this is akin to a Kondo volume collapse (although the system is not in the Kondo regime, and
valence fluctuations are sizeable).
The first-order nature of the transition, already in the purely electronic theory,
implies a tendency towards phase separation, which appears to be present
experimentally.\cite{goko}

This set of ideas holds plausible explanations for
(i) the coincidence of volume-collapse and magnetic--non-magnetic transition
(which is also borne out by first-principles approaches \cite{kreyssig,goldman,antropov2},
(ii) the bad metallic behavior, and
(iii) the reduced magnetic moment of the magnetic phase.
Note that (ii) and (iii) are not described by the mean-field theory, but
arise from residual scattering between itinerant and localized electrons
in the orbital-selective Mott regime as small pressure.
Moreover, the abrupt disappearance of strong magnetism in the high-pressure
phase is a natural part of the story.
Theoretically, superconductivity mediated by residual spin fluctuations emerges at high
pressure and low temperatures. The conflicting experimental reports \cite{kreyssig,tori,yu}
on superconductivity in \cafeas\ may be consistent with the strong sensitivity
of $T_c$ on microscopic parameters expected from theory and with sample inhomogeneities
as proposed in Ref.~\onlinecite{yu}.

The strong first-order volume collapse has only been observed in \cafeas\ --
so what is special about this compound?
While we cannot give a definite answer at this point, the current status of both theory
and experiment suggests that \cafeas\ (i) is a particularly soft material with a small
c-axis lattice constant,
(ii) displays a large electron--phonon coupling, and
(iii) is located in close vicinity to a magnetic--non-magnetic transition.
Indeed, first-principles calculations\cite{yildirim09} have reported a giant
magneto-elastic coupling. In particular, the magnitude of the Fe moment has been found to
be coupled to the c-axis lattice constant, with this effect being strongest in \cafeas\
as compared to other 122 and also to 1111 compounds.
Moreover, a soft lattice is known to be a crucial ingredient to a first-order
volume-collapse scenario.\cite{mahan,allen,cyrot,held,hackl08}
Note that, in our calculations, a small volume jump remains even for hard lattices due to the
first-order nature of the purely electronic transition.

Our conceptual ideas are not in conflict with itinerant spin-density-wave descriptions of the FeAs
magnetism, but constitute a more strong-coupling-inspired view on the same physics. We
speculate that an orbital-selective Mott scenario, likely with a continuous instead of a
first-order transition, could apply to doping-driven transitions in iron arsenides as
well, as a specific filling of the conduction band is not required.


\acknowledgments

We acknowledge useful discussions with
B. B\"uchner, I. Eremin, C. Hess, Y. Qi, A. Rosch, S. Sachdev, and C. Xu.
This research was supported by the DFG through the SFB 608 (K\"oln)
and the Research Units FG 538 and FG 960.



\begin{thebibliography}{99}

\bibitem{kami08}
Y. Kamihara, T. Watanabe, M. Hirano, and H. Hosono,
J. Am. Chem. Soc. {\bf 130}, 3296 (2008).

\bibitem{yang08}
J. Yang, Z.-C. Li, W. Lu, W. Li, X.-L. Shen, Z.- A. Ren, G.-C. Che, X. -L. Dong, L.-L. Sun,
F. Zhou, and Z.-X. Zhao,
Supercond. Sci. Technol. {\bf 21}, 082001 (2008).

\bibitem{taka08}
H. Takahashi, K. Igawa, K. Arii, Y. Kamihara, M. Hirano, and H. Hosono,
Nature {\bf 453}, 376 (2008).

\bibitem{ren08}
Z.-A. Ren, G.-C. Che, X.-L. Dong, J. Yang, W. Lu, W. Yi, X.-L. Shen, Z.-C. Li,  L.-L. Sun, F. Zhou, and Z. Zhao,
Europhys. Lett. {\bf 83}, 17002 (2008).





\bibitem{rotter08}
M. Rotter, M. Tegel, and D. Johrendt, Phys. Rev. Lett. {\bf 101}, 107006 (2008).

\bibitem{buechner08}
H. Luetkens {\em et al.},
Nature Mat. {\bf 8}, 305 (2009).

\bibitem{cruz08}
C. de la Cruz, Q. Huang, J. W. Lynn, J. Li, W. Ratcliff, J. L. Zarestky, H. A. Mook, G. F. Chen, J. L. Luo,
N. L. Wang, and P. Dai,
Nature {\bf 453}, 899 (2008).

\bibitem{klauss08}
H.-H. Klauss {\em et al.},
Phys. Rev. Lett. {\bf 101}, 077005 (2008).

\bibitem{si08}
Q. Si and E. Abrahams,
Phys. Rev. Lett. {\bf 101}, 076401 (2008).

\bibitem{klingeler08}
R. Klingeler, N. Leps, I. Hellmann, A. Popa, C. Hess, A. Kondrat, J. Hamann-Borrero, G. Behr,
V. Kataev, and B. B\"uchner,
preprint arXiv:0808.0708.

\bibitem{lumsden08}
M. D. Lumsden {\em et al.},
Phys. Rev. Lett. {\bf 102}, 107005 (2009).

\bibitem{yildirim08}
Y. Yildirim,
Phys. Rev. Lett. {\bf 101}, 057010 (2008).

\bibitem{mazin08}
I.I . Mazin, M. D. Johannes, L. Boeri, K. Koepernik, and D. J. Singh,
Phys. Rev. B {\bf 78}, 085104 (2008).

\bibitem{kreyssig}
A. Kreyssig {\em et al.},
Phys. Rev. B {\bf 78}, 184517 (2008).

\bibitem{tori}
M. S. Torikachvili, S. L. Budko, N. Ni, and P. C. Canfield,
Phys. Rev. Lett. {\bf 101}, 057006 (2008).

\bibitem{goldman}
A. I. Goldman {\em et al.},
Phys. Rev. B {\bf 79}, 024513 (2009).

\bibitem{yu}
W. Yu, A. A. Aczel, T. J. Williams, S. L. Bud'ko, N. Ni, P. C. Canfield, and G. M. Luke,
preprint arXiv:0811.2554.

\bibitem{pratt}
D. K. Pratt {\em et al.},
Phys. Rev. B {\bf 79}, 060510(R) (2009).

\bibitem{canfield}
P. C. Canfield, S. L. Bud'ko, N. Ni, A. Kreyssig, A. I. Goldman, R. J. McQueeney,
M. S. Torikachvili, D. N. Argyriou, G. Luke, and W. Yu,
preprint arXiv:0901.4672.

\bibitem{antropov2}
G. D. Samolyuk and V. P. Antropov,
Phys. Rev. B {\bf 79}, 052505 (2009).

\bibitem{alireza}
P. L. Alireza {\em et al.}, J. Phys.: Condens. Matter {\bf 21}, 012208 (2009).

\bibitem{phillips}
J. Wu, P. Phillips, and A. H. Castro Neto,
Phys. Rev. Lett. {\bf 101}, 126401 (2008).

\bibitem{dai08}
J. Dai, Q. Si, J.-X. Zhu, and E. Abrahams,
PNAS {\bf 106}, 4118 (2009).

\bibitem{kou}
S.-P. Kou, T. Li, and Z.-Y. Weng,
preprint arXiv:0811.4111.

\bibitem{dai09}
J. Dai, G. Cao, H.-H. Wen, and Z. Xu,
preprint arXiv:0901.2787.

\bibitem{vdb08}
G. Giovannetti, S. Kumar, and J. van den Brink,
Physica B {\bf 403}, 3653 (2008).

\bibitem{laad08}
L. Craco, M. S. Laad, S. Leoni, and H. Rosner,
Phys. Rev. B {\bf 78}, 134511 (2008).

\bibitem{foot_itloc}
Itinerant and local-moment antiferromagnetism are not necessarily distinct,
but can be adiabatically connected as, e.g., in the insulating phase of a single-band
Hubbard model, see:
F. Gebhardt,
{\em The Mott Metal-Insulator Transition--Models and Methods},
Springer, Heidelberg (1997).


\bibitem{mcqueeney}
R. J. McQueeney {\em et al.},
Phys. Rev. Lett. {\bf 101}, 227205 (2008).

\bibitem{mv08}
In heavy-fermion metals, itinerant and local-moment antiferromagnetism
are not qualitatively distinct phases, see:
M. Vojta,
Phys. Rev. B {\bf 78}, 125109 (2008).

\bibitem{foot_other}
For our phenomenological model, the precise orbital character of itinerant and localized
electrons is of little relevance.

\bibitem{hvl}
H. v. L\"ohneysen, A. Rosch, M. Vojta, and P. W\"olfle,
Rev. Mod. Phys. {\bf 79}, 1015 (2007).

\bibitem{coleman}
P.~Coleman, C.~P\'epin, Q.~Si, and R.~Ramazashvili,
J. Phys: Condens. Matt. {\bf 13}, R723 (2001).

\bibitem{edmft}
Q. Si, S. Rabello, K. Ingersent, and J.~L. Smith,
Nature (London) \textbf{413}, 804 (2001);
Phys. Rev. B \textbf{68}, 115103 (2003).

\bibitem{senthil}
T. Senthil, S. Sachdev, and M. Vojta, Phys. Rev. Lett. \textbf{69}, 216403 (2003).

\bibitem{senthil2}
T. Senthil, M. Vojta, and S. Sachdev, Phys. Rev. B \textbf{69}, 035111 (2004).

\bibitem{pepin}
C. Pepin, Phys. Rev. Lett. {\bf 98}, 206401 (2007).

\bibitem{hackl08}
A. Hackl and M. Voj\-ta,
Phys. Rev. B {\bf 77}, 134439 (2008).

\bibitem{schmalian}
M. Dzero, M. R. Norman, I. Paul, C. P\'epin, and J. Schmalian,
Phys. Rev. Lett. \textbf{97}, 185701 (2006).

\bibitem{landau}
L. D. Landau and I. M. Lifshitz, { \em Theory of Elasticity, Theoretical
Physics,} vol. 7, Butterworth-Heinemann, Oxford (1986).

\bibitem{mahan} A. K. McMahan, C. Hushcroft, R. T. Scalettar, and E. L. Pollock,  J. of Comp.-Aided Mat. Design
\textbf{5}, 131 (1998).

\bibitem{allen}
J. W. Allen and R. M. Martin, Phys. Rev. Lett. \textbf{49}, 1106 (1982).

\bibitem{cyrot}
M. Lavagna, C. Lacroix, and M. Cyrot, Phys. Lett. \textbf{90A}, 210 (1982);
J. Phys. F \textbf{13}, 1007 (1983).



\bibitem{held}
K. Held, A. K. McMahan, and R. T. Scalettar, Phys. Rev. Lett. \textbf{87}, 276404 (2001);
A. K. McMahan, K. Held, and R. T. Scalettar, Phys. Rev. B \textbf{67}, 075108 (2003).


\bibitem{osmott1}
V. Anisimov, I. Nekrasov, D. Kondakov, T. Rice, and
M. Sigrist, Eur. Phys. J. B {\bf 25}, 191 (2002);
S. Biermann, L. de Medici, and A. Georges,
Phys. Rev. Lett. {\bf 95}, 206401 (2005).

\bibitem{j1j2a}
P. Chandra and B. Doucot,
Phys. Rev. B {\bf 38}, 9335 (1988).

\bibitem{j1j2b}
J. Oitmaa and W. H. Zheng,
Phys. Rev. B {\bf 54}, 3022 (1996).



\bibitem{kiv_nematic}
C. Fang, H. Yao, W.-F. Tsai, J.-P. Hu, and S. A. Kivelson,
Phys. Rev. B {\bf 77}, 224509 (2008).


\bibitem{xu08}
Y. Qi and C. Xu,
preprint arXiv:0812.0016.

\bibitem{antropov}
K. D. Belashchenko and V. P. Antropov,
Phys. Rev. B {\bf 78}, 212515 (2008).

\bibitem{gammafoot}
The strain-induced changes of $J_{1,2}$ are likely dominated by changes in the bond
angles.\cite{yildirim08}

\bibitem{burdin}
S. Burdin, D. R. Grempel, and A. Georges, Phys. Rev. B \textbf{66}, 045111 (2002).

\bibitem{coqblin}
B. Coqblin, C. Lacroix, M. A. Gusmao, and J. R. Iglesias,
Phys. Rev. B {\bf 67}, 064417 (2003).

\bibitem{saddlefoot}
For our parameter values,
we have checked that saddle points with magnetic order at $(\pi,\pi)$
as well as those with $s$-wave-like pairing
have a higher free energy.

\bibitem{haule}
J. H. Shim, K. Haule, and G. Kotliar,
preprint arXiv:0809.0041.

\bibitem{xiang}
F. Ma, Z. Lu, and T. Xiang,
preprint arXiv:0806.3526.

\bibitem{andersen}
A. N. Yaresko, G.-Q. Liu, V. N. Antonov, and O. K. Andersen,
Phys. Rev. B {\bf 79}, 144421 (2009).

\bibitem{goko}
T. Goko {\em et al.},
Phys. Rev. B {\bf 79}, 144421 (2009).

\bibitem{yildirim09}
T. Yildirim,
Phys. Rev. Lett. {\bf 102}, 037003 (2009);
preprint arXiv:0902.3462.
\end{thebibliography}
\end{document}